\documentclass[]{revtex4}
\usepackage{amssymb,epsfig,amscd}
\usepackage{graphicx}
\usepackage{dcolumn}
\usepackage{bm}

\usepackage{epsfig}

\def\be{\begin{equation}}
\def\ee{\end{equation}}
\def\bq{\begin{eqnarray}}
\def\eq{\end{eqnarray}}

\begin{document}

\title{Time-reversal odd distribution functions in chiral 
models with vector mesons} 
\author{Alessandro Drago}
\address{Dipartimento di Fisica, Universit{\`a} di Ferrara and
INFN, Sezione di Ferrara, 44100 Ferrara, Italy}


\begin{abstract}
The so-called time-reversal odd distribution functions
are known to be non-vanishing in QCD due to the presence
of the link operator in the definition of these quantities.
I show that T-odd distributions can be
non-vanishing also in chiral models, if vector mesons are
introduced as dynamical gauge bosons of a hidden local symmetry. 
Moreover, since the flavor dependence of these distributions
is different in chiral models respect to non chiral ones, 
the phenomenological analysis
of experimental data will be able to distinguish between
these two classes of models.
\end{abstract}


\noindent

\maketitle
\section{Introduction}
The history of time-reversal odd distribution functions is 
interesting and rather complicated. Apparently, 
the first of these quantities
has been initially introduced by Sivers 
to explain single-spin asymmetries \cite{Sivers:1989cc,Sivers:1990fh}
and later used in several phenomenological analysis by
Anselmino, Boglione and Murgia \cite{Anselmino:1994tv,Anselmino:1998yz}.
Another T-odd distribution was introduced by Boer, Mulders and 
Tangerman and it was used to explain the angular dependence of 
the lepton pair in
unpolarized Drell-Yan \cite{Mulders:1995dh,Boer:1997nt,Boer:1999mm}.
Both these quantities were considered to be forbidden in QCD since they
apparently change sign under the combined parity and time-reversal
transformation \cite{Collins:1992kk}. Their theoretical status
remained therefore uncertain till the discovery by
Brodsky, Hwang and Schimdt of an explicit mechanism which
could originate single-spin asymmetries in QCD \cite{Brodsky:2002cx}.
There, a gluon exchange between the struck quark and the
target spectator was able to generate a phase difference
between this amplitude and the
lowest order contribution. Moreover, the existence of a 
non-vanishing transverse momentum $k_\perp$ allows the asymmetry to be
not power-law suppressed in $Q^2$ ,as long as $k_\perp$ is small
compared to $Q$.  
Collins recognized that the presence of the link operator,
preserving the gauge invariance of the matrix element, was 
sufficient to make ``T-odd'' distribution functions non-vanishing in
QCD \cite{Collins:2002kn}
and, immediately after, Ji, Yuan and Belitsky 
\cite{Ji:2002aa,Belitsky:2002sm}
clarified how the
expansion of the link operator could explicitly provide a contribution
of the type discussed in Ref.~\cite{Brodsky:2002cx}.
More recently, a few model calculations appeared of the 
Sivers function \cite{Yuan:2003wk,Bacchetta:2003rz,Lu:2004au},
all of them making use of the gluon exchange
mechanism of Ref.~\cite{Brodsky:2002cx} in order to generate the 
phase needed for a non-vanishing T-odd distribution.

An interesting question concerns the possibility of computing
T-odd distributions using chiral models, in which quarks 
are interacting via
the exchange of chiral fields instead of gluons. 
The main problem here is that apparently the link operator
is not well defined, since chiral models
are in general not considered to be gauge theories and therefore no
gauge field is present. One could be tempted to naively
substitute the chiral fields to the gluon field in the link operator,
but this is clearly arbitrary, since the need for the link operator
in the definition of the T-odd distribution is strictly connected 
with a local gauge symmetry, which is apparently absent in 
chiral models. On the other hand, one can remark that
it would be rather surprising that in chiral models
T-odd distribution functions are identically zero, since
it would constitute, at least to my knowledge, the first
example of a quantity, not directly involving gluons, 
which, while non-vanishing in QCD, cannot be estimated 
using a chiral lagrangian. 
This impasse can actually be circumvented if one
recall that in the 80s vector mesons have been introduced
in chiral lagrangians following two approaches, which
at the end can be shown to be equivalent. In the first scheme, 
a hidden local
symmetry is shown to be present in chiral lagrangians 
\cite{Bando:1984ej,Bando:1984pw,Meissner:1986ka,Bando:1987br,Meissner:1987ge}
and vector mesons are the gauge fields of this local symmetry.
In the second approach, vector mesons are introduced as massive
Yang-Mills fields of the chiral $\mathrm{U}(N_f)_L\otimes\mathrm{U}(N_f)_R$
symmetry \cite{Kaymakcalan:1983qq,Gomm:1984at,Kaymakcalan:1984bz}. 
For simplicity, I will adopt the first scheme, although
also the second one should lead to the same result. 
The main aim of the papers written in the 80s was to provide a model for 
vector meson phenomenology. Here, I will mainly be interested
in the formal aspect of these models, in which a  
local gauge invariance is present, in order to show that 
T-odd distributions can be non-vanishing also in chiral models.

\section{The hidden symmetry}
The idea of a hidden local symmetry was proposed in
supergravity theories \cite{Cremmer:1978ds,Cremmer:1979up} and it 
states that any nonlinear sigma model based on the manifold
$G/H$ is gauge equivalent to another model with
$G_\mathrm{global}\otimes H_\mathrm{local}$ symmetry and 
the gauge bosons of the hidden local symmetry $H_\mathrm{local}$
are composite fields 
\cite{Golo:1978de,D'Adda:1978uc,D'Adda:1978kp,Breitenlohner:1984rr}.
In hadronic physics the nonlinear sigma model
$\mathrm{SU(2)}_L\otimes\mathrm{SU(2)}_R/\mathrm{SU(2)}_V$
is extensively used as a model for low energy phenomenology
and the diagonal subgroup is here $H=\mathrm{SU(2)}_V$. 
In Ref.\cite{Bando:1984ej} the $\rho$ meson was therefore
interpreted as the gauge boson of the local hidden group $H$,
and in \cite{Meissner:1986ka} the gauge symmetry was extended
to the $\mathrm{SU(2)}_V\otimes \mathrm{U(1)}$ group,
incorporating the phenomenology
of the $\omega$ meson in this scheme.

To recall what the hidden symmetry in chiral models
is, let us start
from the kinetic term of the lagrangian of the nonlinear sigma model, 
following Ref.\cite{Bando:1984ej}:
\be
L=\left (f_\pi^2/4\right )
\mathrm{Tr}\left (\partial_\mu U\partial ^\mu U^\dagger \right)\, ,\label{lagr}
\ee
where 
\be
U(x)=\exp[2 i \pi(x)/f_\pi]
\ee
with $\pi\equiv \pi^a T^a$. Here $T^a$ are the generators of the
$\mathrm{SU(2)}$ group and $f_\pi$ is the pion decay constant.
Under the global symmetry the field transforms as 
\be
U(x)\rightarrow g_L U(x) g_R^\dagger \, ,
\ee
where $g_L$ and $g_R$ are elements of $\mathrm{SU(2)}_L$ and 
$\mathrm{SU(2)}_R$, respectively.
It is now possible to rewrite the lagrangian (\ref{lagr}) so that
it will exhibits a local symmetry, besides the global one. 
We rewrite the field $U(x)$ in terms of two variables, 
$\xi_L(x)$ and $\xi_R(x)$, as:
\be
U(x)\equiv\xi_L^\dagger(x)\xi_R(x)
\ee
and we introduce the gauge field $V_\mu\equiv V_\mu^a(x) T^a$. 
Finally, we define the transformation rules of these fields
under the group
$[\mathrm{SU(2)}_L\otimes\mathrm{SU(2)}_R]_\mathrm{global}
\otimes [\mathrm{SU(2)}_V]_\mathrm{local}$ as:
\bq
\xi_L(x)&\rightarrow& h(x)\xi_L(x) g_L^\dagger \nonumber \\
\xi_R(x)&\rightarrow& h(x)\xi_R(x) g_R^\dagger\\ 
V_\mu(x)&\rightarrow& ih(x)\partial_\mu h(x)^\dagger+h(x)V_\mu(x)h(x)^\dagger
\nonumber \, .
\eq
We can now define a covariant derivative as:
\be
D_\mu\xi_{L,R}(x)\equiv[\partial_\mu-iV_\mu(x)]\xi_{L,R}(x)\, ,
\ee
and we can recognize that it is possible to write two invariants
under $[\mathrm{SU(2)}_L\otimes\mathrm{SU(2)}_R]_\mathrm{global}
\otimes [\mathrm{SU(2)}_V]_\mathrm{local}$, namely:
\bq
L_V&=&-\left (f_\pi^2/4\right )\mathrm{Tr}\left (D_\mu\xi_L\cdot\xi_L^\dagger+
        D_\mu\xi_R\cdot\xi_R^\dagger\right )^2 \label{lv}\\
L_A&=&-\left (f_\pi^2/4\right )\mathrm{Tr}\left (D_\mu\xi_L\cdot\xi_L^\dagger-
        D_\mu\xi_R\cdot\xi_R^\dagger\right )^2\, .\label{la}
\eq
Any linear combination of (\ref{lv}) 
and (\ref{la}) is equivalent to the original lagrangian (\ref{lagr}),
as it is easy to check choosing e.g. the gauge
$\xi_L^\dagger=\xi_r\equiv\xi=\exp(-i \pi f_\pi)$, for which
$L_A=L$ and $L_V=0$.

As already mentioned, 
the previous analysis was extended in Ref.~\cite{Meissner:1986ka}
so that the $\omega$ meson can also be interpreted as a gauge
field.

Up to now no dynamics has been attributed to the gauge field
$V_\mu$ and it can therefore be eliminated in terms of the chiral fields
by solving field equations. The reason to introduce a dynamics
associated with the gauge field is twofold. From one side it is 
possible to obtain a very successful description of the vector meson
phenomenology by identifying the gauge field with the $\rho$ meson. Moreover,
one can speculate that the kinetic term is originated by quantum
effects. 
Finally, the mass of the vector mesons is obtained via 
the Higgs mechanism, in which the unphysical scalar modes
in $\xi_{L,R}$ are absorbed into the $\rho$ meson.
For simplicity, I will also assume the existence of a kinetic term
for the vector mesons, although this is not a necessary condition
in order to get non-vanishing T-odd distributions, but in this 
way I will be able to easily reproduce the result obtained in
Ref.~\cite{Brodsky:2002cx} where a massive photon field
was assumed to be exchanged between the struck quark and the
spectator diquark.

\section{Introducing quarks}

The hidden symmetry mechanism was first introduced in 
models as the Skyrme one, in which only chiral fields
are present. It is anyway possible, and relatively
straightforward, to extend this idea including the 
quark sector 
(see e.g.~\cite{Bando:1984pw,Simic:1985jx,Meissner:1987ge,RuizArriola:1990am}).
In this way we can obtain models which have a more direct
contact with QCD and can be used to evaluate matrix
elements of quark operators as, e.g., quark distribution functions.
Following in particular Ref.~\cite{Bando:1984pw}, quarks can
be introduced assuming that they transform as a fundamental
representation of $H_\mathrm{local}$ and as singlet of 
$G_\mathrm{global}$ (the so-called
``constituent gauge'' of Ref.~\cite{Simic:1985jx}).
The quark kinetic term 
\be
L_q=\bar\psi i \gamma^\mu\partial_\mu\psi
\ee
can be made invariant under the group
$[\mathrm{SU(2)}_V]_\mathrm{local}$,
under which quarks transform as
\be
\psi(x)\rightarrow h(x) \psi\, ,
\ee
by introducing the covariant derivative already defined
in the previous section 
\be
D_\mu\psi(x)\equiv[\partial_\mu-iV_\mu(x)]\psi(x)\, .
\ee
In this way, an ``effective QCD'' lagrangian
can be defined as
\be
L_\mathrm{eff}=L_A+a L_V+\bar\psi i \gamma^\mu D_\mu\psi +...
\ee
in which vector mesons are introduced as the gauge fields
of a local gauge transformation. 

As in any gauge theory, when dealing with bilocal operator
a link operator need to be introduced in order to preserve
local gauge invariance. 
As usual, the link is defined to be:
\be
W(x_1,x_2)\equiv 
\mathrm{P.O.}\,\exp\,\left(i\int_{x_1}^{x_2} \mathrm d s_\mu V^\mu\right)\, .
\ee
Therefore, when considering e.g.
a quark distribution function, the gauge invariance of the 
theory requires the existence of the link operator. At this point
the contact with QCD is re-established and the matrix element
defining the partonic distribution can be evaluated within the
effective lagrangian in the same way as it can be computed in QCD. 

\section{T-odd distributions}

We can now return to the original question, namely the 
possibility of computing T-odd distributions using chiral models.
The argument of Collins \cite{Collins:2002kn} is that the 
definition of the distribution functions requires the existence
of the link operator to preserve $\mathrm{SU(3)}_c$ gauge
invariance. When deep-inelastic scattering is considered, 
the factorization scheme dictates the use of
future-pointing Wilson lines, while for Drell-Yan the Wilson
lines are past-pointing. Therefore 
under time-reversal the distribution functions appropriate
for Drell-Yan transform into the distribution functions
appropriate for DIS, but for a sign.
It is clear that  
the problem of the existence of T-odd distributions in QCD
is then solved, because the definition itself of these quantities
explicitly contains a specific direction in time. Therefore, no
problem connected with the breaking of time reversal symmetry exists,
as long as a link operator can be unambiguously defined. 
This is precisely what chiral lagrangian with vector mesons
can provide, as shown in the previous section, if 
vector mesons are introduced as gauge fields of a local symmetry.
 
To be more explicit, let me remark that the calculation of
the Sivers function provided in \cite{Brodsky:2002cx}, was actually
based on the exchange of a massive photon instead of a gluon,
and the mass was introduced to eliminate infrared divergences.
When using vector-meson exchange instead of gluons the mass of
the mesons provides a natural infrared regulator. But for that,
the calculation proceeds strictly parallel to the one of
Ref.~\cite{Brodsky:2002cx} and a non-vanishing Sivers distribution
can therefore be obtained.

\section{Conclusions}
I will now discuss the phenomenological implications of the 
possibility of computing T-odd distributions, and particularly the
Sivers function, using chiral models. This possibility was discussed
in the past using sigma models
\cite{Anselmino:1996qm,Anselmino:1997jj,Anselmino:2001vn,Anselmino:2002yx},
but unfortunately in those papers no link operator was introduced and
therefore the correct way to circumvent 
the time-reversal problem was not found. 
It is nevertheless interesting to remark
that a phenomenological relation was derived, showing that,
at leading order in the $1/N_c$ expansion the Sivers function for
the up quark is equal and opposite to the Sivers function for the
down quark:
\be
\Delta_0^T f_u(x,\mathbf k_\perp)=
-\Delta_0^T f_d(x,\mathbf k_\perp)\, . \label{sim}
\ee
This same relation was later obtained in a 
model independent way by Pobylitsa \cite{Pobylitsa:2003ty},
unfortunately also there without indicating the origin of 
the link operator. In this paper I have provided an
example of a chiral model in which the link operator is clearly defined.
Since the possibility of introducing vector mesons in chiral models
is largely independent on the details of the model at hand, one can
conclude that relation (\ref{sim}) can indeed be obtained in
those models, with both sides of the relation non-vanishing.
More explicitly, the main difference between the calculation
of the Sivers function in chiral models and e.g. the calculation
of Ref.~\cite{Brodsky:2002cx} is in the wave function of the proton
which provides the spin decomposition. In chiral models the spin
structure is obtained as an expansion in $1/N_c$ and from there
the relation (\ref{sim}) can be derived in a rather direct way 
\cite{Pobylitsa:2003ty}.
It is interesting to notice that in non-chiral models the 
Sivers function for the down quarks turns out to be much smaller than that
for the up quarks \cite{Bacchetta:2003rz,Lu:2004au}, 
opening the possibility to discriminate
between various model previsions using already collected data
and the ones obtainable in future experiments 
\cite{Airapetian:2004tw,Webb:2005cd,Pagano:2005jx,PAX,Anselmino:2005nn}.

\bigskip
It is a pleasure to thank M.~Anselmino, D.~Comelli and 
P.~Ferretti Dalpiaz for many
useful discussions.

\bibliography{references}
\bibliographystyle{apsrev}

\end{document}